\documentclass[twocolumn,aps,prd,longbibliography]{revtex4-2}
\usepackage{amsmath,latexsym}
\usepackage{xcolor}
\usepackage[colorlinks=true, urlcolor=blue, linkcolor=blue, citecolor=blue]{hyperref}
\usepackage{etoolbox}
\usepackage{breqn}
\usepackage{multirow}
\usepackage{tipa}
\usepackage{lipsum}
\usepackage{float}
\usepackage{booktabs}
\usepackage{makecell}
\DeclareUnicodeCharacter{0301}{\'{e}}
\makeatletter
\let\cat@comma@active\@empty
\makeatother

\begin{document}
\preprint{APS/123-QED}

\title{Dynamics and frictional dissipation from treading in the puddle}

\author{Chung-Hao Chen, Zong-Rou Jiang,  and Tzay-Ming Hong\thanks{ming@phys.nthu.edu.tw} }
\thanks{ming@phys.nthu.edu.tw}
\affiliation{Department of Physics, National Tsing Hua University, Hsinchu, Taiwan 30013, Republic of China}

\date{\today}

\begin{abstract}
It was recently established that dogs share the same lapping technique as cats by flicking their tongue against the water surface and then yanking it back, dragging up a column of water. This liquid column appears frequently in daily life and industrial applications, such as walking through a puddle and roller printing. While governed by the Navier-Stokes equation, its dynamics are often studied by numerical means, which hinders a full understanding of the rich mixture of physics behind, for instance, the competition of surface and potential energies, and how the pinch-off is affected by the kinetic energy and water jet when a large cylinder is used.
Combined with simple models, we elucidate the mechanism that drives the change of morphology and derive analytic expressions for the critical height and upper radius for the liquid column when transiting between three stages. Stage I is characterized by a static and reversible profile for the column whose upper radius $r_t$ equals that of the cylinder. The column becomes irreversible and $r_t$ starts shrinking upon entering stage II.
It is not until $r_t$ stops shrinking that the column neck accelerates its contraction and descends toward the pool, the quantitative behavior of which is among the successful predictions of our theory. Pinch-off dominates the second half of stage III without its usual signature of self-similarity. This is discussed and explained with an interesting incident involving a water jet similar to that made by a dropping stone.
\end{abstract}

\maketitle


\section{\label{sec:level1} Introduction}
We often encounter puddles when venturing outside on rainy days. The liquid column that forms below and then detaches from our shoes is omnipresent in many daily phenomena, such as the lapping of a cat \cite{catdrinkwater} and dog \cite{dogdrinkwater}, the wondrous feat of walking on water by the Jesus lizard \cite{jesus_lizard}, roller printing \cite{Roller_printing}, capillary feeders \cite{Prakash_2008}, micro-robots \cite{micro_robots} and crystal growth \cite{crystal_growth}.
This is why the lifting process of a liquid column has attracted academic and industrial interest. One practical application is the inject printer which involves the ejection of an ink column from a nozzle. The uniformity of ink delivered by each droplet formed after the pinch-off due to Rayleigh-Plateau instability concerns the quality of the printer. The physics behind these phenomena centers on the force created when water molecules cling together, i.e., the surface tension.
One standard device to study this process involves dipping a cylinder into a liquid pool, and then pulling it up. 

Observing the water column's evolution as we lift our shoes above the puddle makes it easy to notice that its profile is static and reversible initially. Only after reaching a critical height $H^\star$ does it start to contract by itself, implying a transition from being reversible to irreversible. To determine the static properties, one has to employ the Young-Laplace equation \cite{Lord_Kelvin_1886,Blaisdell_1940,Staicopolus_1962,Staicopolus_1967}. Plateau \cite{Plateau_1863} is the first person to use the Young-Laplace equation to discuss the stability condition of a liquid column in 1863. A decade later, Howe \cite{Howe_1887} extended the stability condition to the unbounded case via the variational method without gravitational effect and assumed volume to be a constant. It was not until 1971 that Padday \cite{1971_Padday} 
incorporated gravity effects and systematically examined different varieties of boundary conditions. Built on these pioneering works, more researchers joined the effort by studying more complex systems, such as when multiple liquid columns were connected \cite{1984_A.V.Zhdanov}, different shapes of capillary bridges \cite{Mason_1970,Gillette_1971,Sanz_1983,Russo_1986,Slobozhanin_1993,Meseguer_1985,Perales_1991,Meseguer_1984,Slobozhanin_1998}, and when the length of the liquid column becomes longer than its circumference \cite{1994_B.J._Lowry}.

Instead of solving the Young-Laplace equation which is mathematically complex, insights were gained by an alternative view that emphasized the importance of examining the free energy change during the pulling process. This new approach was inspired by the observation of Dombrowski and Franzer \cite{1954_Dombrowski} that only those holes that exceeded the critical radius  $r_{c}$ would grow in size during the disintegration of liquid sheets
produced from the single-hole fan-spray nozzle and the spinning disk. Almost two decades later, Taylor and Michael \cite{1972_G.I._Taylor} noticed disturbances on the liquid sheet from the recording video, and attributed them to the close-up of small holes. Treating the hole as an analogy of the liquid column with the role of air and liquid reversed, the sheet thickness is equivalent to the column height. To verify their idea, the authors theoretically calculated and experimentally examined the relation between $r_{c}$ and the sheet thickness on solid surfaces with varying contact angles by considering the change of free energy. They also found that $r_c$ went to infinity when the sheet exceeded some critical thickness. Ashutosh and Ruckenstein \cite{1989_Ashutosh}  
corrected the critical thickness theoretically by incorporating the solid-liquid interfacial tension in 1989.
Debregea and Brochard-Wyar \cite{Nucleation_radius} confirmed in 1997 that the conclusion of Ref.\cite{1972_G.I._Taylor} applied to the liquid column, i.e.,  the column will either grow or shrink, depending on whether its radius is greater or smaller than  $r_c$ which value increases monotonically with $H$.
 The perturbation method can also be used to solve the problem \cite{Vega_1983}.

Interesting complex dynamics ensues soon until the liquid column becomes irreversible. Theoretical efforts have been made by 
the time-dependent formulation \cite{1985_Jose} and the implicit numerical methods \cite{1986_I._Martinez}.  
Detailed dynamics leading up to the eventual breakup of a capillary surface were discussed by Cryer and Steen in 1992 \cite{1992_Steven_A}. 
Theoretical calculations now have to rely on solving the Navier-Stokes equation numerically.
Due to the complexity of mathematical equations and lack of high-speed recording technique in the '90s, the contact line at the top of the liquid column was always assumed to be pinned during the pulling process \cite{Gaudet_1996,Zhang_1996,Ambravaneswaran_1999,Yildirim_2001,Kang_2009,Dodds_2009,fix_boundary_example_1}. However, the boundary does move before pinch-off, which effect was not taken into account until 1999 \cite{2002_C.A._Powell,2011_Bian_Qian,2012_Shawn_Dodds,Christodoulou_1997,1999_Ruben_Scardovelli,Troian_2001,Sethian_2003,2015_Bian_Qian,2018_Jyun_Ting_Wu,Richard_2000,P_dimi,2005_M.A.Walkley,2007_Xueying_Xie,2010_Zhou,soft_matter,static_2013}.

Even though the statics and dynamics for the liquid column can be reproduced by numerically solving the Young-Laplace and Navier-Stokes equations, the goal of this manuscript is to combine experimental studies with minimal models that enable (1) more detailed identification of different regions that often elude our naked eyes and (2) analytic expressions for the critical parameters that characterize their transition.

\section{Experimental Setup and results}

\begin{figure}[htbp]
    \centering
    \includegraphics[scale=0.35]{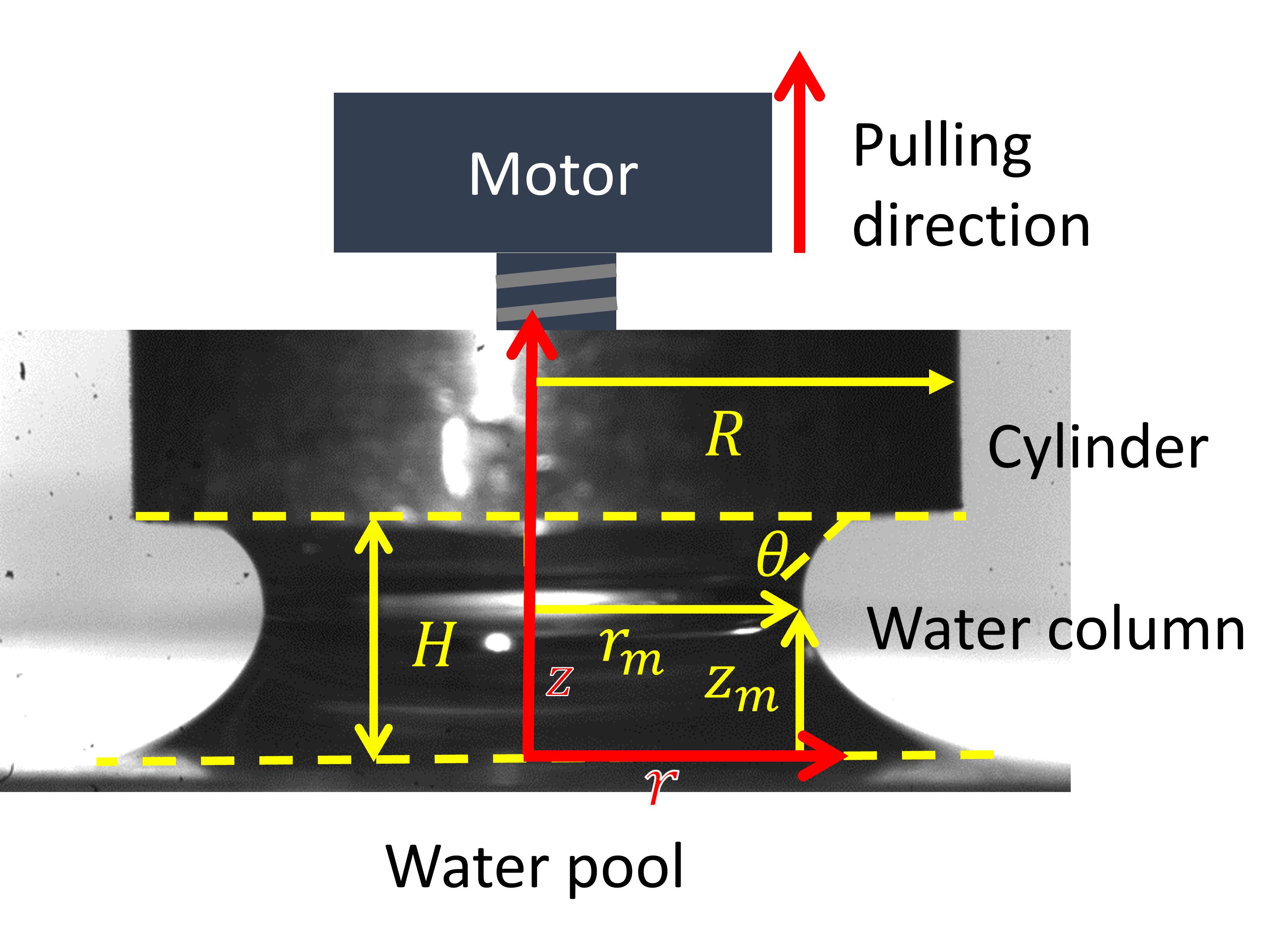}
    \caption{A water column of height $H$ is generated when an aluminum cylinder is pulled out of the pool. 
    While $\theta$ and $r_t$ denote the contact angle and radius at the top of the water column, $r_{ m}$ and $z_{m}$ are the radius and height at its neck. }
    \label{fig:enter-label}
\end{figure}

We use aluminum cylinders of 14 different diameters ranging from 1 to 26 mm. Initially dipped in the water pool, the cylinder is raised at a constant speed as slow as 0.5 mm/s  by a stepper motor to form a water column when pulled above the pool surface, as shown in Fig. \ref{fig:enter-label}. The levelness of the cylinder is checked by the level gauge attached at its top. 
In the meantime, we employ a high-speed camera with 8000 fps to capture the evolution of the column profile.

\begin{figure}[htbp]
    \centering
    \setlength{\leftskip}{-10pt}
    \includegraphics[scale=0.3]{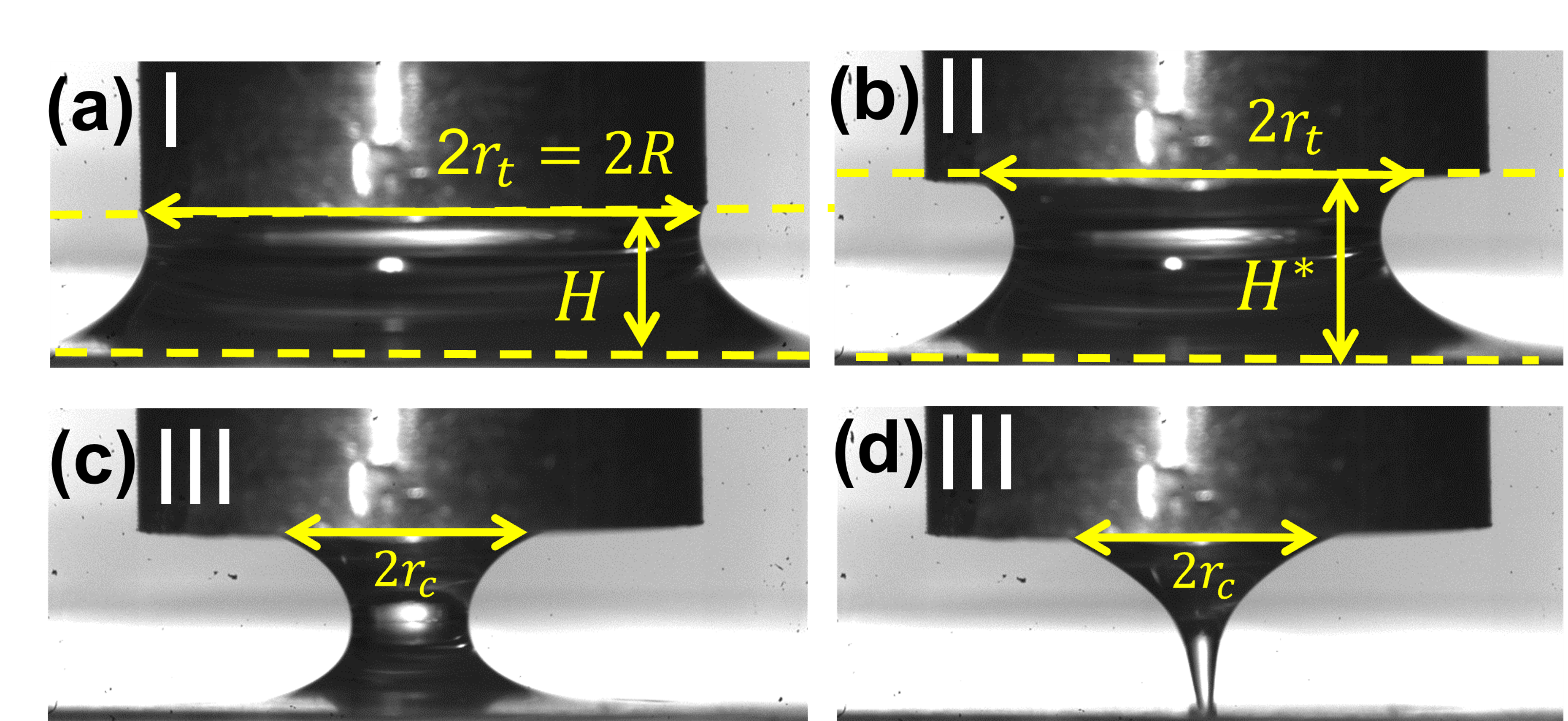}
    \caption{ The profile of the water column goes through three stages: (a) $r_t$ remains fixed at $R$ as  $H$ is raised, (b) $r_t$ starts shrinking after $H$ reaches and is kept at $H^\star$ - last for about 0.1$\sim$0.3 s, and (c, d) $r_t$ appears to stay at $r_c$, while the column proceeds with the pinch-off - last for about 0.004$\sim$0.006 s.}
\label{evolve}
\end{figure}

\begin{figure}
    \centering
    \setlength{\leftskip}{-5pt}
    \includegraphics[scale=0.4]{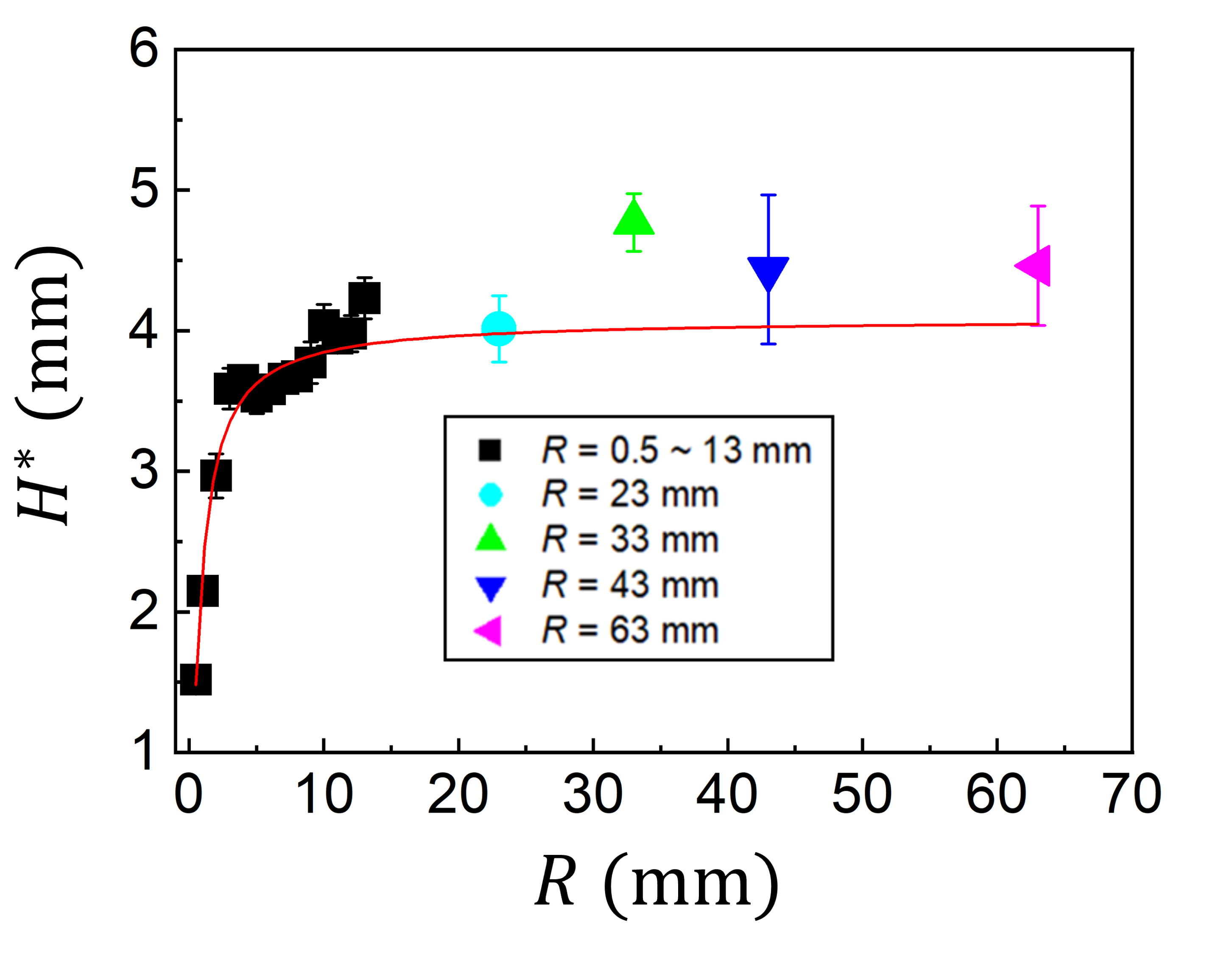}
    \caption{The $H^{\star }$ is plotted against $R$. The red solid line represents the theoretical prediction of Eq. (\ref{Hstar}). }
    \label{H_star_vs_R}
\end{figure}

\begin{figure}
    \centering
    \setlength{\leftskip}{-20pt}
    \includegraphics[scale=0.38]{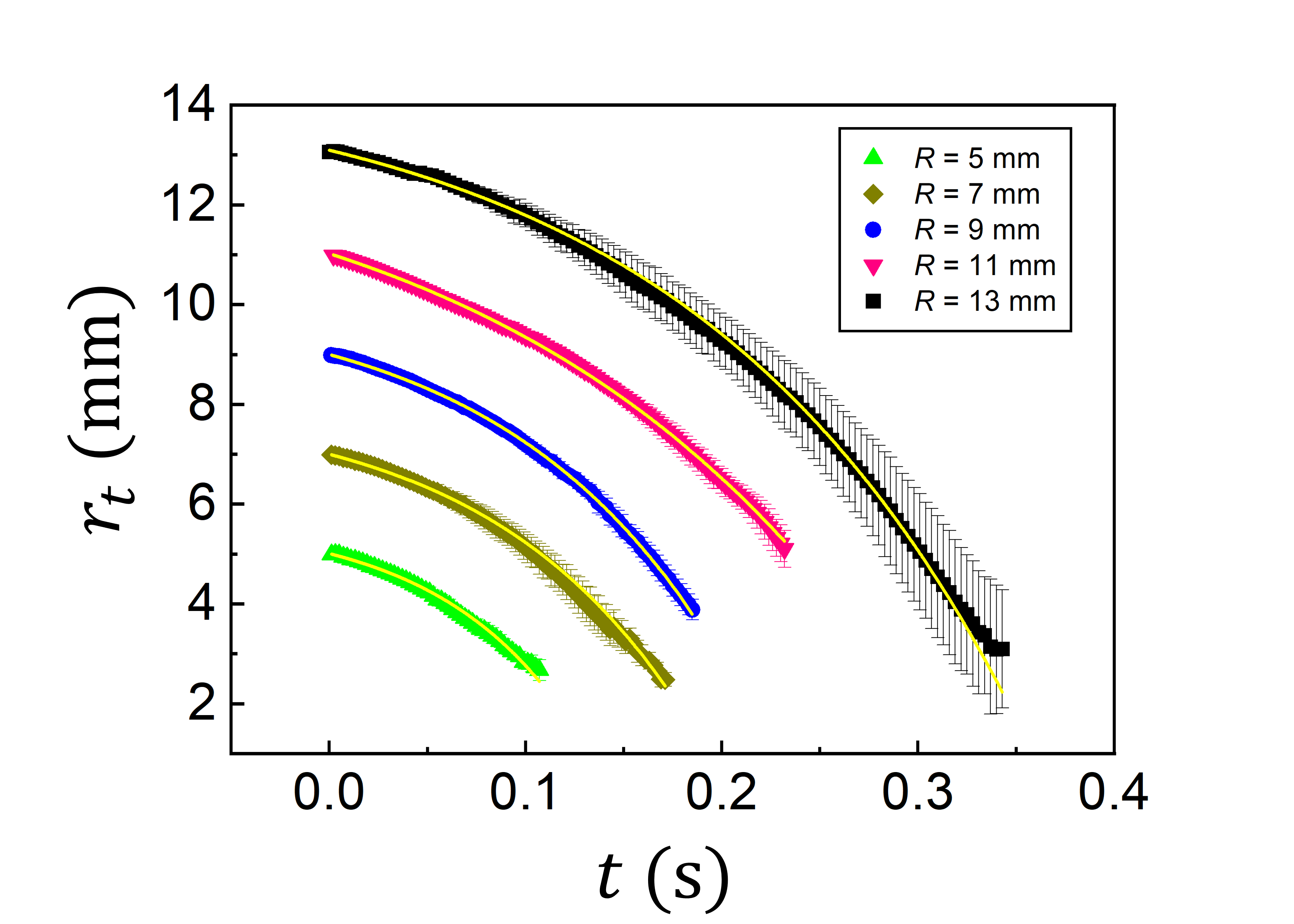}
    \caption{The  $r_t$ shrinks with $t$ in stage II for  $R$=5, 7, 9, 11, and 13 mm. The yellow solid lines are predictions by Eq. \eqref{rt}.}
    \label{R_top vs t}
\end{figure}

As shown in Fig. \ref{evolve}, the profile of the water column pulled by a cylinder of radius $R\leq 13$ mm undergoes three stages from being reversible in (a) to being irreversible in (b) and (c, d), which we shall denote by stages I, II, and III. The critical height $H^\star$ that marks the transition from stage I to II is plotted as a function of $R$ in Fig. \ref{H_star_vs_R} that appears to level off at large $R$. We also record how the top radius $r_t$ of the column shrinks with time $t$ in stage II in Fig. \ref{R_top vs t} for five different $R$. 

\begin{figure}[h]
    \centering
    \setlength{\leftskip}{-25pt}
\includegraphics[scale=0.35]{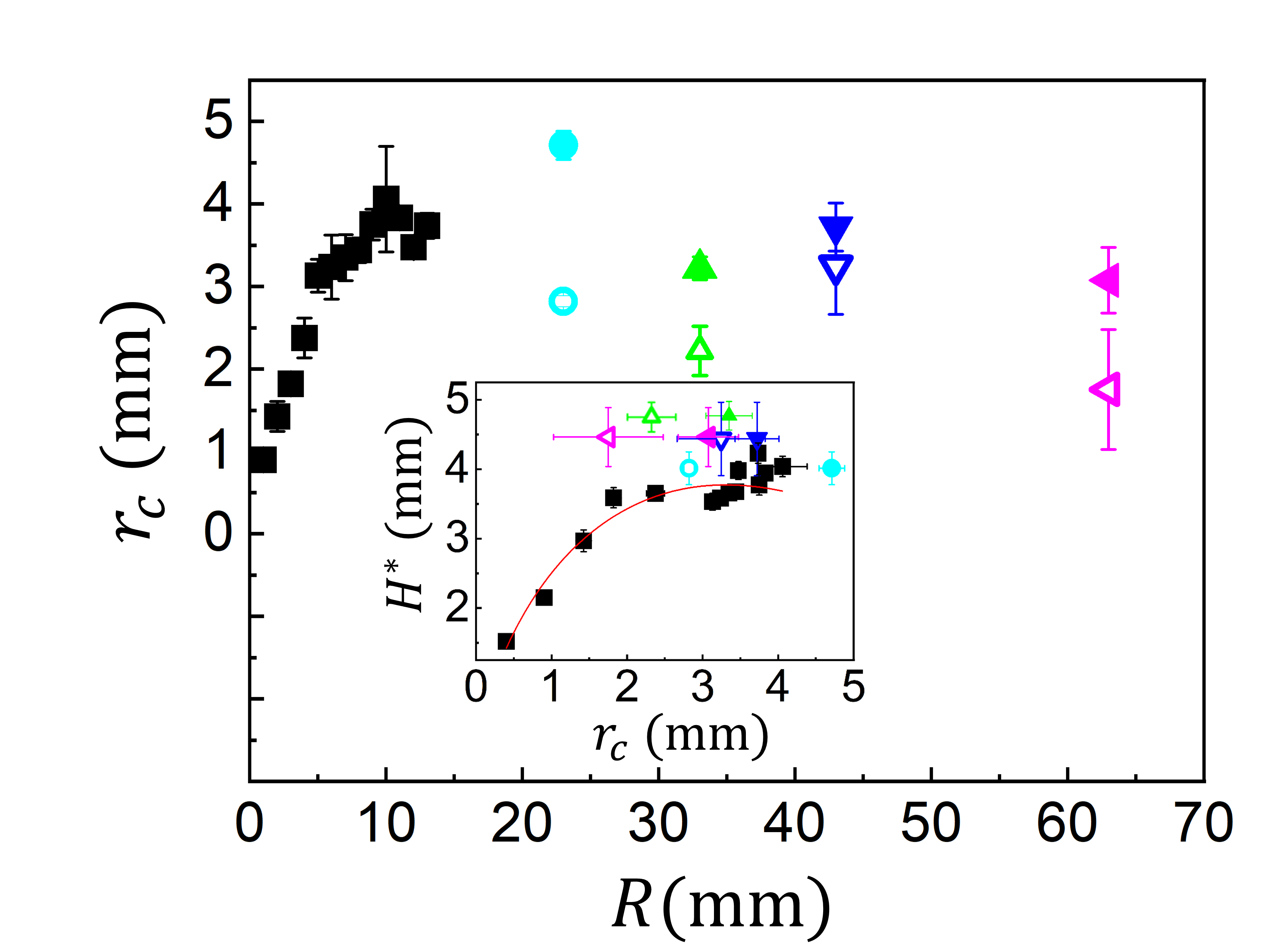}
    \caption{ The $r_c$ is plotted against $R$ where the labeling follows that of Fig. \ref{H_star_vs_R}. The red solid fitting line comes from plugging Eq. \eqref{Hstar} into Eq. \eqref{rc}. A bump appears at roughly $R$=13 mm.  
    The inset shows the relation between $H^\star$ and $r_{c}$ that matches well the prediction of Eq.  (\ref{rc}) in the red solid line, except for $R>$ 23 mm when the residual kinetic energy of liquid column shrinks $r_c$ further. The hollow colored labels represent $r_{c2}$. }
    \label{rc_vs_R}
\end{figure}

The critical radius $r_c$ at which $r_t$ saturates toward the end of stage II follows the same trend as  $H^\star$, i.e., increases with $R$ initially in Fig. \ref{rc_vs_R}. However, the trend is reversed when $R$ exceeds roughly 13 mm, hinting at the occurrence of new events. An indication was glimpsed in stage III when a bulge of water visibly climbed up the column from the pool and interrupted the pinch-off by spreading open the neck. This visible clump of water then bounces off from the cylinder and plunges into the pool. Afterward, the delayed process of pinch-off resumes. 


\section{ Theoretical models }
\subsection{\label{sec:level2}Stage I: reversible}
\begin{figure}
    \centering
    \includegraphics[scale=0.35]{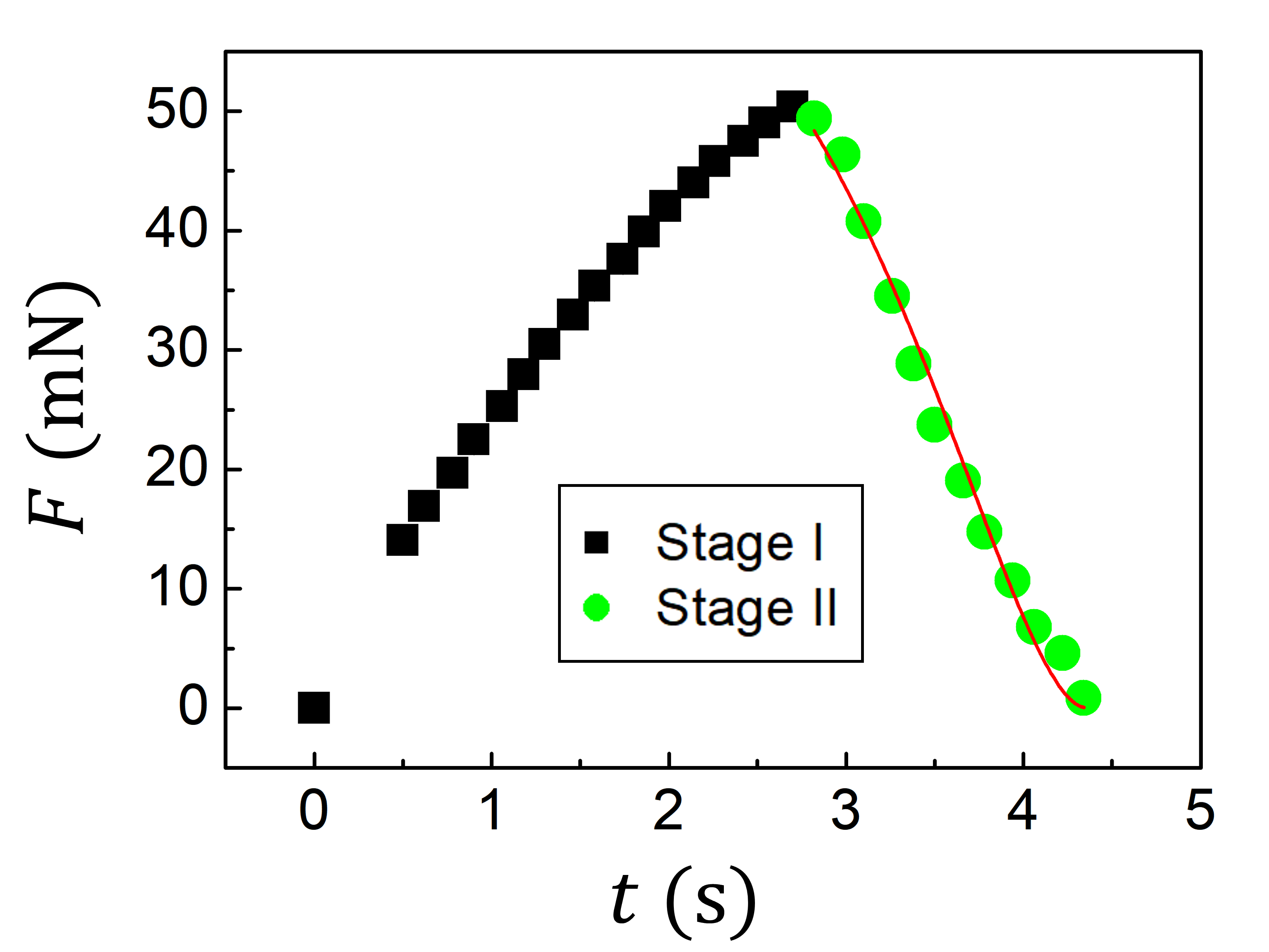}
    \caption{The pull force vs. time for $R$ = 23 mm. Each data point comes from the average of five sets of experiments. The linear growth in stage I and the decay in stage II are consistent with our prediction from Eqs. \eqref{total} and \eqref{rt}. Note that the force for stage III and the error bars are too small to show.  }
 \label{force}
\end{figure}

An analytic expression of $H^{\star }$ is possible by optimizing the total energy of the system with respect to $r_{t}$: 
\begin{equation}
    E_{\rm total}=2\pi r_{t}H\sigma_{lg}+\pi r^{2}_{t}\rho g \frac{H^{2}}{2}+(\sigma_{ls}-\sigma_{gs}-\sigma_{lg})\pi
    r^{2}_{t}
\label{total}
\end{equation}
where the water column has been approximated by a uniform cylinder, $\rho$ denotes the mass density of water, and $\sigma_{lg}$, $\sigma_{ls}$, and $\sigma_{gs}$ are the surface tension constant for liquid-vapor, liquid-solid, and vapor-solid interfaces. 
The force $F$ required to pull the cylinder can be obtained by differentiating Eq.  \eqref{total} to $H$. The predicted linear relation matches the data in Fig. \ref{force}.

\begin{figure}
    \centering
    \setlength{\leftskip}{0pt}
    \includegraphics[scale=0.38]{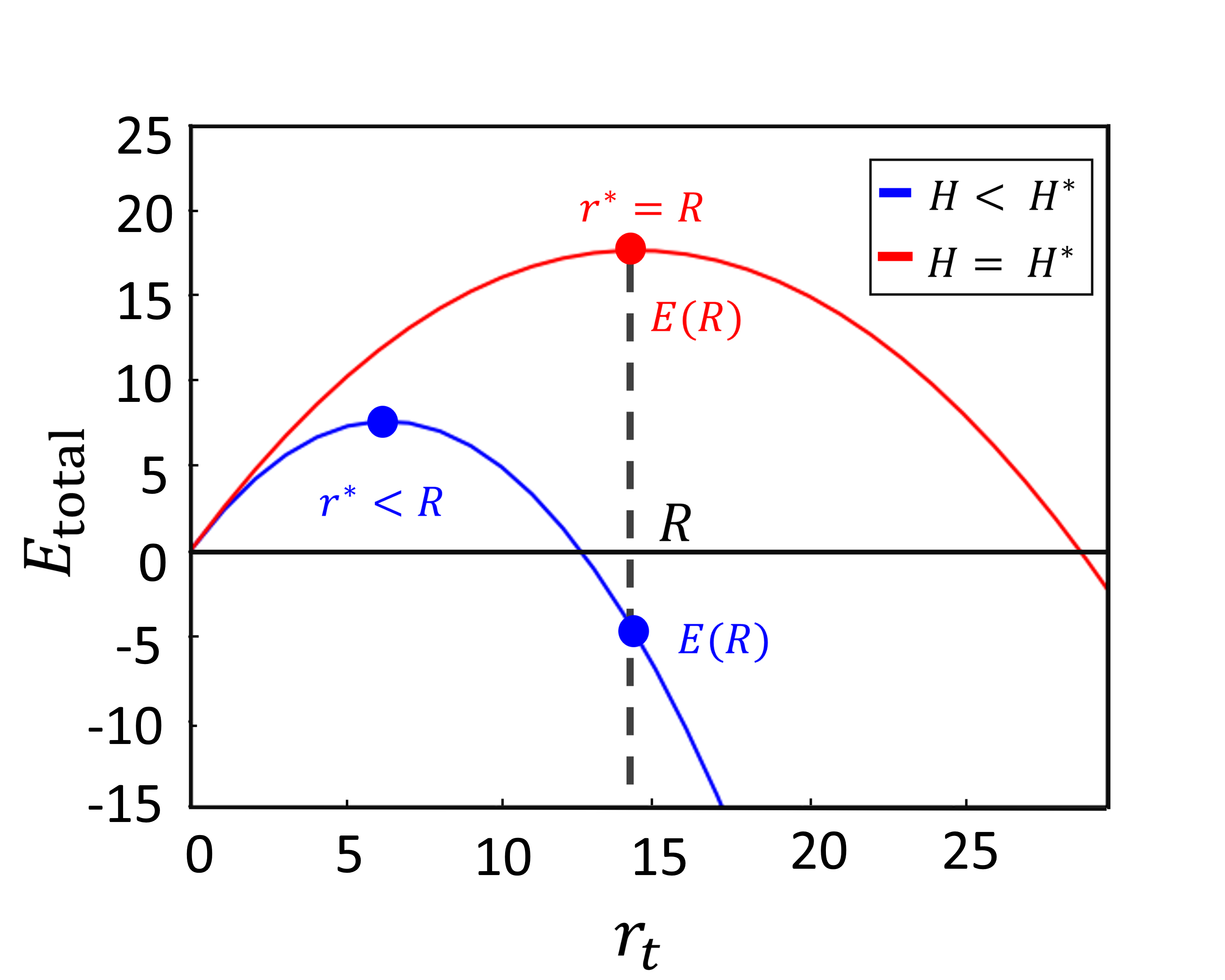}
    \caption{Equation (\ref{total}) is numerically plotted against $r_{t}$. When $H<H^\star$, it is energetically unfavorable for $r_t$ to shrink from $R$, as is evident from the blue line. 
    As $H$ reaches $H^{\star }$ and the system switches to obeying the red line, $r_t=R$ becomes an energy maximum, and any perturbation will tip the balance and trigger an irreversible shrinkage. Note that the possibility of expansion is ruled out since $r_t$ cannot exceed $R$.}
    \label{H_star theory}
\end{figure}

We know from Young's formula that $\Delta \sigma\equiv \sigma_{gs}+\sigma_{lg}-\sigma_{ls}$ is positive definite. When plotted against $H$, we define the maximum of Eq.  (\ref{total}) occurs at $r_t= r^\star$:
\begin{equation}
     r^{\star }=\frac{H\sigma_{lg}}{\Delta\sigma-\frac{\rho g H^{ 2}}{2}}.
\label{rstar}
\end{equation}
  If $R>r^\star$, $r_{t}$ will remain at $R$ where $E_{\rm total}$ is at its local minimum,  as shown by the blue line in Fig. \ref{H_star theory}.
As we raise $H$, $r^\star$ increases according to Eq.  \eqref{rstar}. As soon as $r^\star =R$, $r_t=R$ flips from a local minimum to a maximum which triggers an irreversible shrinkage of $r_t$.
Therefore, $H^\star$ can be obtained by setting  $r^\star =R$ in Eq.   \eqref{rstar}:
 \begin{equation}
    H^{\star } =  \frac{\sigma_{lg}\Big[\sqrt{1+\frac{2\rho g R^{2}\Delta \sigma}{\sigma_{lg}^{2}}}-1\Big]}{\rho g R}.
 \label{Hstar}
 \end{equation} 
This prediction fits the experimental data in Fig. \ref{H_star_vs_R}  with R-square = 0.96 and explains why $H^\star$ saturates at large $R$.

\subsection{Stage II: irreversible and quasi-static}

Once $H$ reaches $H^{\star }$, $r_{t}$ starts to shrink voluntarily. The shrinkage rate can be modeled by
\begin{equation}
     \frac{dE_{\rm total}}{dr_t}= -A\frac{dr_t}{dt}
\end{equation}
where $A$ is a phenomenological constant. Solving this differential equation gives 
\begin{equation}
    r_t(t) = \frac{\alpha}{2\beta} -Be^{\frac{2\beta }{A} t}
\label{rt}
\end{equation}
where $\alpha$ and $-\beta$ correspond to the coefficients of $r_t$ and $r_t^2$ in Eq. (\ref{total}), and the constant $B$ signifies a perturbation to tip the balance, as mentioned in Fig. \ref{H_star theory}. Equation (\ref{rt}) agrees well with the experimental data in Fig. \ref{R_top vs t}.
Since $H^\star$ is held constant in stage II, the pull force $F=\rho \pi r_t^{2}H^{\star}$ can be obtained from the weight of water column and Eq.  (\ref{rt}). This prediction matches the data in Fig. \ref{force}. 

\begin{figure}[htbp]
    \setlength{\leftskip}{-10pt}
    \includegraphics[scale=0.34]{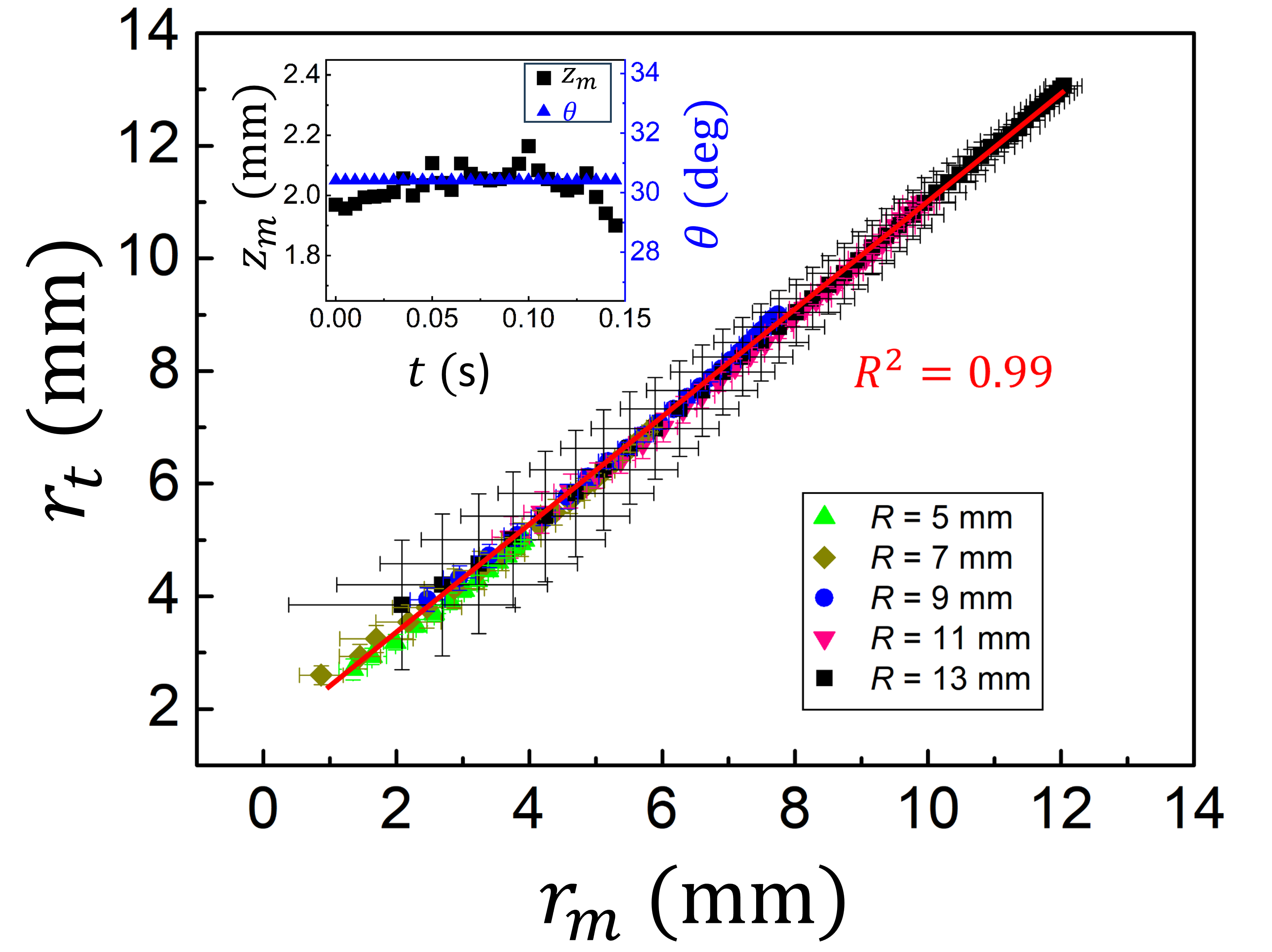}
    \caption{Both $r_{t}$ and $r_{m}$ shrink with time in regime II, but their values appear to be linearly proportional, as verified by the red solid fitting line from Eq. (\ref{r_m vs r_t}). Inset shows that $z_m$ and $\theta$ are not sensitive to time. }    
    \label{R_top vs R_min}
\end{figure}

Although stage II is irreversible, we argue that it is quasi-static based on two observations: first, the shape of the water column remains unchanged because $r_{t}$ and  $r_{m}$ exhibit the same shrinking speed and their correlation is strongly linear in Fig. \ref{R_top vs R_min}. Second, the contact angle $\theta$ and $z_m$ defined in Fig. \ref{fig:enter-label} are not affected by the receding motion, as shown in the inset of Fig. \ref{R_top vs R_min}. As a result, we can assume the profile or $r(z)$ is still dictated by the minimization of its surface energy and gravitational energy at each step of the way:
\begin{equation}
    E = \sigma_{lg}\int_{0}^{H^{\star }}2\pi r \sqrt{1+r^{'2}}\ dz +\int_{0}^{H^{\star }}\rho g \pi r^{2}z\ dz
\end{equation}
By use of the Euler-Lagrange equation, we obtain
\begin{equation}
    \frac{\sigma_{lg}}{\rho g}\Big[\frac{1}{\sqrt{1+r^{'2}}}-\frac{rr^{''}}{(\sqrt{1+r^{'2}})^{3}}\Big]+rz=0
    \label{euler}
\end{equation}
which cannot be solved analytically. To obtain an approximate solution, we appeal to the
constraints on $r'$ that (a) $r' \rightarrow -\infty$ as $z\rightarrow 0$, (b) $r'\rightarrow 0$ as  $z\rightarrow z_{m}$, and (c) $r'=\cot{\theta}$ as  $z=H^{\star }$.
The simplest guess that satisfies all these conditions is that
\begin{equation}
\label{whole slope}
    r'=\frac{z^{2}-z_{m}^{2}}{z}\frac{H^{\star }}{H^{\star 2}-z_{m}^{2}}\cot\theta .
\end{equation}
Our choice of $(z^{2}-z_{m}^{2})$ over $(z-z_{m})$ to satisfy constraint (b) is based on setting  $z\rightarrow z_{m}$ in  Eq. (\ref{euler}) to obtain
\begin{equation}
    \frac{\sigma_{lg}}{\rho g}(1-rr'')+rz=0.
\label{temp}
\end{equation}
From Fig.  \ref{evolve}(b), we know  $rr''\approx 20\gg 1$ so that Eq. (\ref{evolve}) can be further simplified to give  
\begin{equation}
\label{z->zm slope}
    r'=\frac{\rho g}{\sigma_{lg}}(\frac{z^{2}-z_{m}^{2}}{2}).
\end{equation}
Comparing the coefficients of Eqs.(\ref{whole slope})  and (\ref{z->zm slope}) gives
\begin{equation}
    \frac{\rho g}{2\sigma_{lg}}=\frac{H^{\star }{\cot\theta}}{z_{m}(H^{\star 2}-z_{m}^{2})}
\label{13}
\end{equation}
that uniquely determines $z_m$ as a function of 
 $\theta$, $H^{\star }$ and $\sigma_{lg}$. This is consistent with the empirical finding that  $z_m$ is roughly independent of time in the inset of Fig. \ref{R_top vs R_min}. 

Finally, after solving Eq. (\ref{whole slope}) with the boundary condition $r(z=H^\star )=r_t$, we set  $z=z_m$ to obtain:
\begin{equation}
    r_{t}=r_{m}+\Big[\frac{1}{2}-
    \frac{z_{m}^{2}}{H^{\star 2}-z_{m}^{2}}\ln(\frac{H^{\star }}{z_{m}})\Big]H^{\star }\cot\theta
    \label{r_m vs r_t}
\end{equation}
Since $z_{m}$ is a constant from Eq. (\ref{13}), the linear relation between $r_{t}$ and $r_{m}$ in Fig.  \ref{R_top vs R_min} is derived.

\subsection{Stage III: irreversible and pinch-off}

Since stage III lasts for only about 0.01 s, $r_{t}$ appears to be stationary at $r_c$. Why and how the Rayleigh-Plateau instability is triggered at this stage needs to be answered.  We know that the pull from the surface tension, $2\pi r_t \sigma_{lg} \sin{\theta}$, is not big enough to sustain the weight of water in stage II, which is why the column has to keep shedding its load. Since the former is proportional to $r_t$, while the latter is roughly $r_t^2$ when $H^\star$ is fixed, the surface tension soon catches up with the weight as $r_t$ shrinks. We conjecture that they become equal at $r_t=r_c$, and the momentum to shrink that is carried over from stage II sets off the pinch-off phenomenon.

Although developing a singular neck in stage III is a dynamic process, $r_c$ is predetermined at its borderline with stage II where a quasi-static approximation is feasible. So we can still perform the variational method on the potential energy to find how $r_c$ is decided by $H^\star$: 

\begin{equation}
    E = \sigma_{lg} \int_{0}^{H^{\star}} 2\pi r\sqrt{1+{r'}^2} dz + \lambda \int_{0}^{H^{\star}} \pi {r}^{2}dz
\end{equation}
where the surface energy is assumed to dominate the gravitational energy since the volume of the water column has decreased by more than fourfold since stage I. The Lagrange multiplier is incorporated to make sure that the water volume equals
\begin{equation}
    \int_{0}^{H^{\star}}\pi r^{2}dz=\frac{2\pi r_c \sigma_{lg} \sin{\theta}}{\rho g}.
    \label{vol}
\end{equation}
Using the second form of the Euler-Lagrange equation, we obtain
\begin{equation}
\label{Euler lagrange Rc}
    \frac{r}{\sqrt{1+{r'}^{2}}}+\frac{\lambda r^2}{2\sigma_{lg}} = C
\end{equation}
where the constant $C$ must equal $r_c\sin{\theta}+\lambda r_c^{2}/2\sigma_{lg}$, $r_{m}+\lambda r_{m}^{2}/2\sigma_{lg}$,
and  $\lambda r(0)^{2}/2\sigma_{lg}$ at the same time for $z=H^\star$, $z_m$, and 0, respectively. Since the pool radius $r(0)$ can be much bigger than $r_c$ and $r_m$, the second terms of the first two expressions are much smaller than the third expression. Therefore, we know that \begin{equation}
r_c\sin\theta\approx r_m\approx 
\frac{\lambda r(0)^{2}}{2\sigma_{lg}}.
\label{trick}
\end{equation}

Plug the expression of $r'$ from Eq. (\ref{Euler lagrange Rc}) in $z=\int dr/r'$ and set $z=H^\star$ to obtain
\begin{equation}
\label{Hc integral}
   \frac{H^{\star }}{r_c}= \Big[\int_{\frac{r_{m}}{r_c}}^{1}+\int_{\frac{r_{m}}{r_c}}^{\frac{r(0)}{r_c}}\Big]\frac{dx}{\sqrt{\big[\frac{x}{\sin{\theta}+\frac{\lambda r_c }{2\sigma_{lg}}(1-x^2)}\big]^{2}-1}} 
\end{equation}
which is to be solved in conjunction with Eq. (\ref{vol}):
\begin{equation}
\label{volume conservation}
\frac{2\sigma_{lg}\sin{\theta}}{\rho g r_c}=
\Big[\int_{\frac{r_{m}}{r_c}}^{1}+\int_{\frac{r_{m}}{r_c}}^{\frac{r(0)}{r_c}}\Big]
\frac{x^{2}dx}{\sqrt{\big[\frac{x}{\sin{\theta}+\frac{\lambda r_c }{2\sigma_{lg}}(1-x^2)}\big]^{2}-1}}
\end{equation}
where the factor $\frac{\lambda r_c }{2\sigma_{lg}}=\frac{{r_c}^2 \sin\theta}{r^2(0)}\ll 1$ from Eq. (\ref{trick}).
Notice that both equations are dominated by the upper limit, $x_0=\frac{r(0)}{r_c}\gg 1$, of their second integral near which 
\begin{equation}
\frac{1}{\sqrt{\big[\frac{x}{\sin{\theta}+\frac{\lambda r_c }{2\sigma_{lg}}(1-x^2)}\big]^{2}-1}}\approx 
\frac{\sin{\theta}-\frac{\lambda r_c }{2\sigma_{lg}}x^2}{x}
\end{equation}
where $\frac{\lambda r_c }{2\sigma_{lg}}x_0^2=\sin\theta$.
This greatly simplifies the integration to give $\sin\theta \ln x_0$ and $\sin\theta x_0^2/4$ for Eqs.(\ref{Hc integral})  and (\ref{volume conservation}), respectively.
Combining these two results immediately predicts that
\begin{equation}
    \frac{H^{\star }}{r_c} \approx \frac{\sin{\theta}}{2}\ln{\frac{8\sigma_{lg}}{\rho g r_c}}.
\label{rc}
\end{equation}
which agrees well with the experimental data in the inset of Fig. \ref{rc_vs_R}.

\begin{figure}
    \includegraphics[scale=0.35]{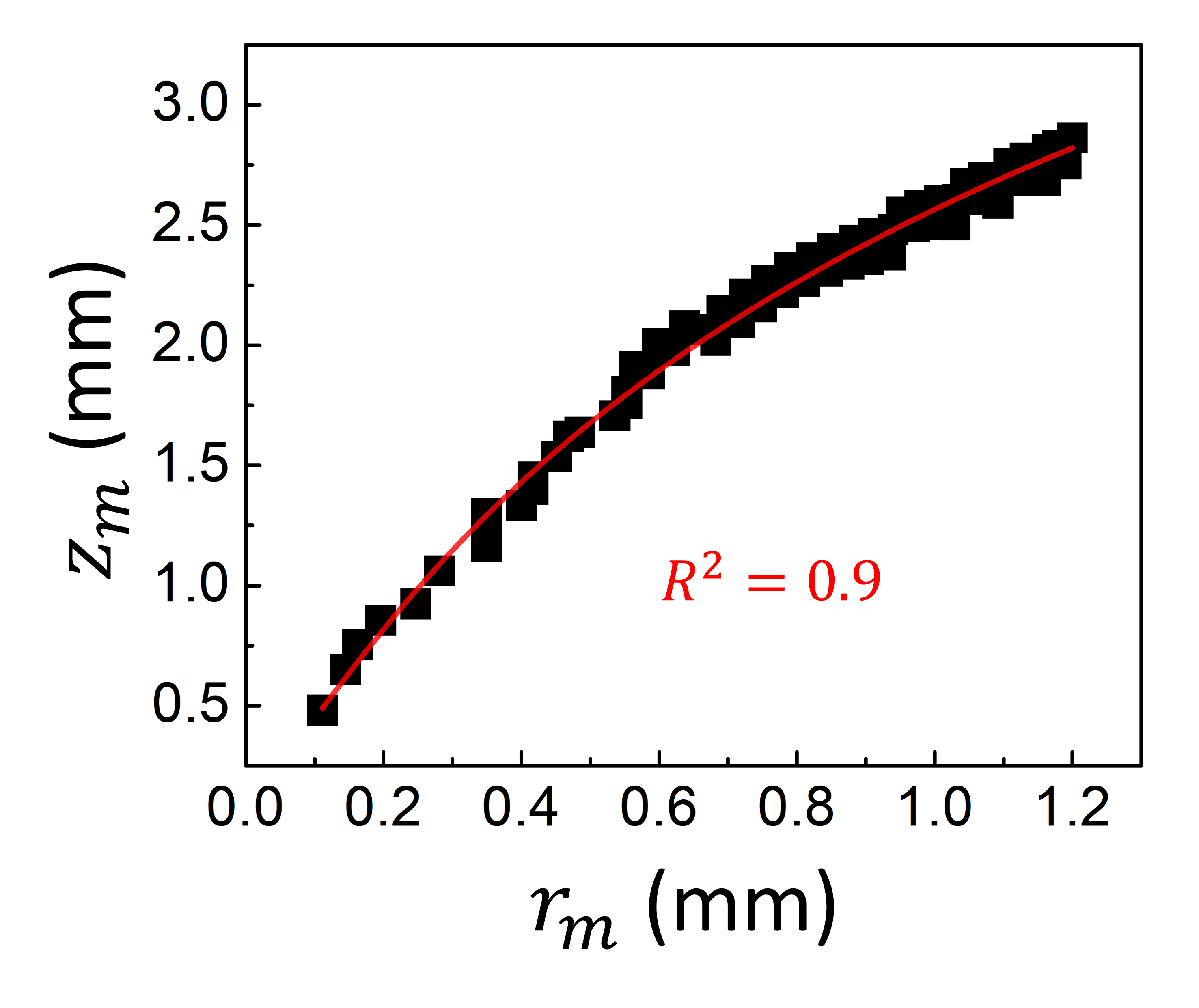}
    \caption{Relation between $z_{m}$ and $r_{m}$ in stage III for $R$=4 mm. The red solid fitting line is based on Eq. (\ref{r_m vs r_t}).}
    \label{zm vs rm}
\end{figure}

\begin{figure}[ht]
    \centering
    \setlength{\leftskip}{-10pt}
    \includegraphics[scale=0.55]{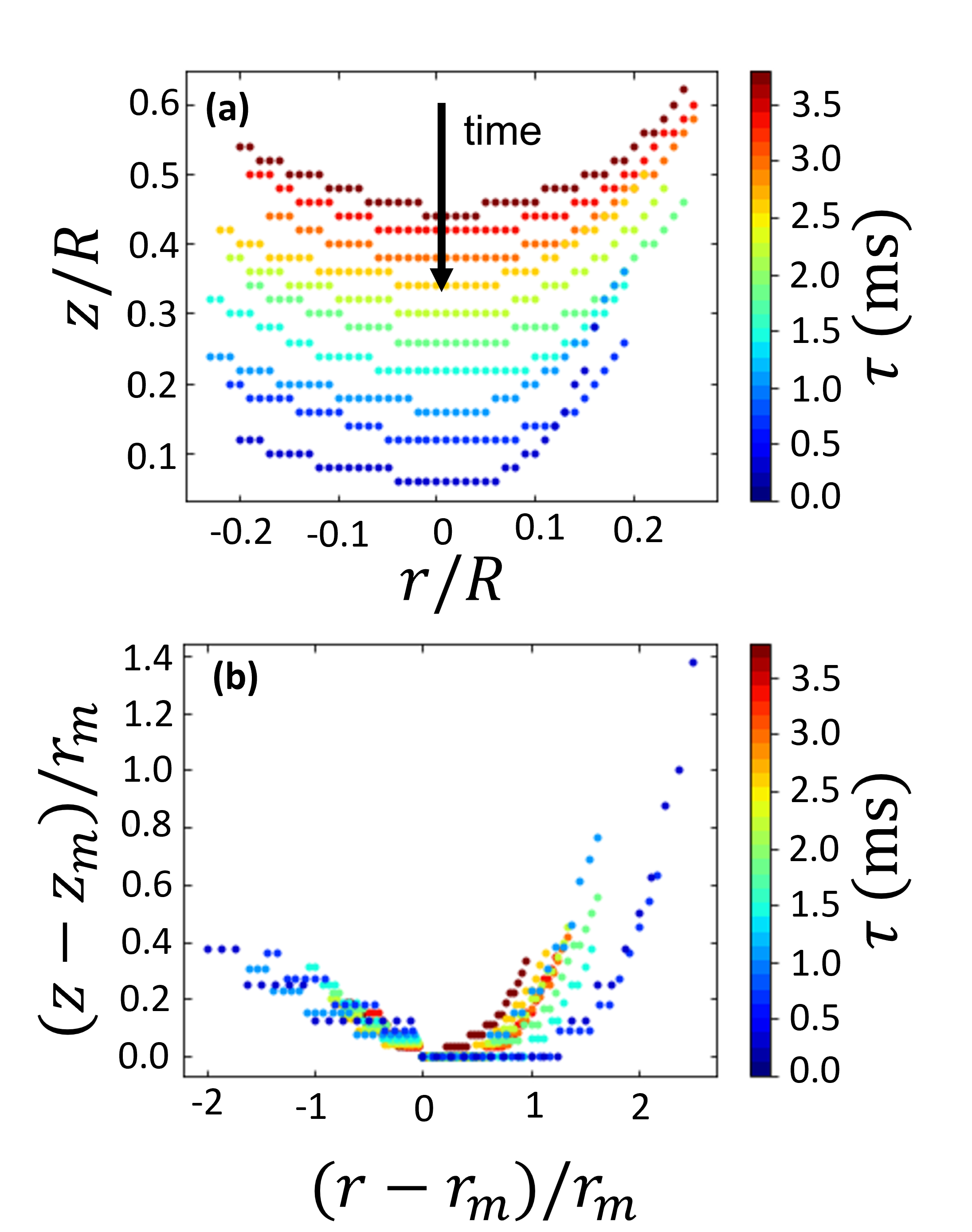}
    \caption{(a) The evolution of profile $z(r)$ in the vicinity of column neck is shown during stage III. Color bars quantify the remaining time. (b) No self-similarity is found after rescaling \cite{rescale} plot (a). }
 \label{similar}
\end{figure}

\begin{figure}[ht]
    \centering
    \setlength{\leftskip}{-10pt}
    \includegraphics[scale=0.35]{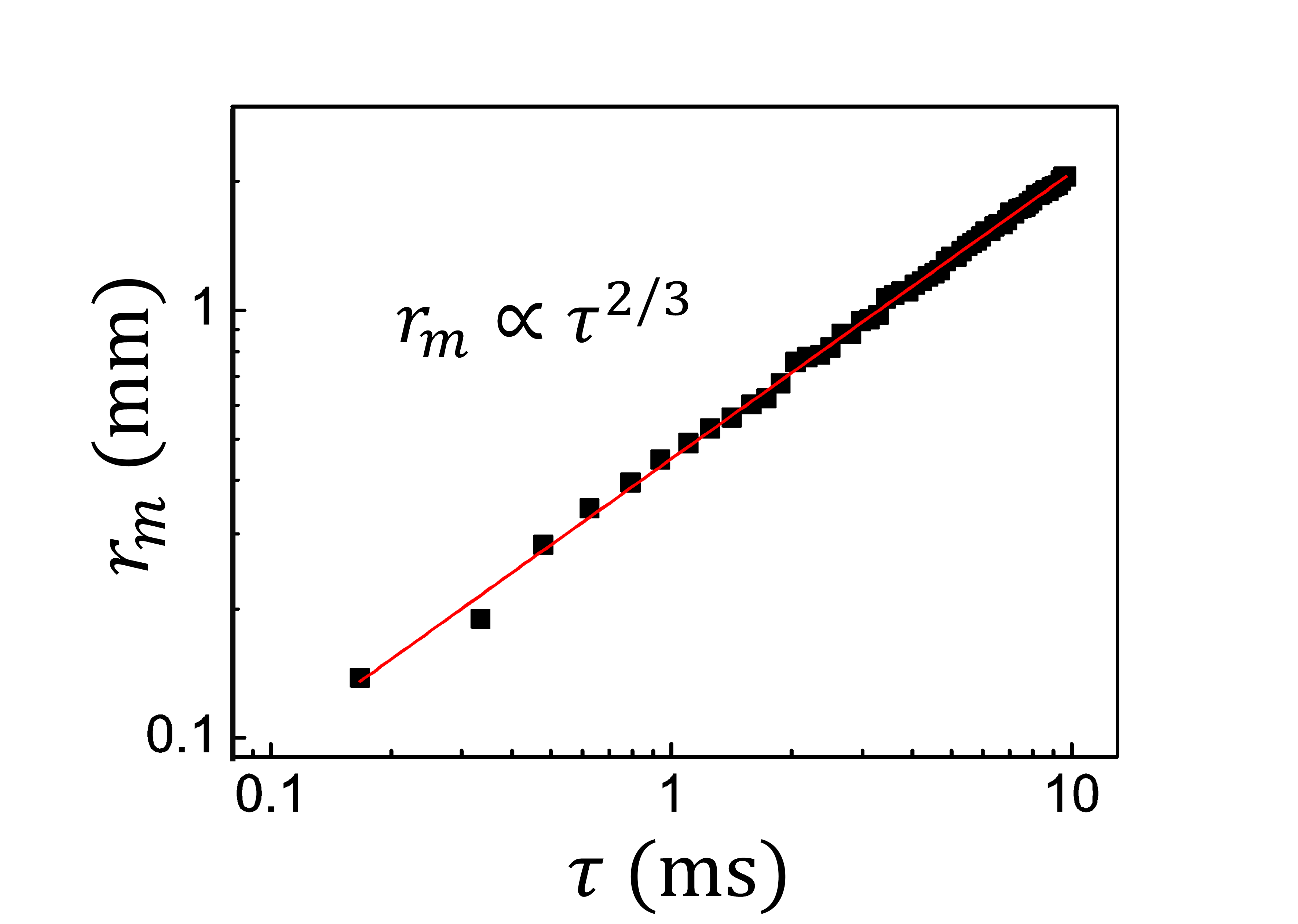}
    \caption{Log-log plot for $r_{m}$ vs the  remaining time $\tau = t_{p}-t$ until pinch-off, where $t_{p}$ is the time of pinch-off. The scaling exponent is 2/3.}
    \label{scalingrelation}
\end{figure}

During the pinch-off, the bottleneck plunges toward the pool surface before breaking up, as shown in Fig. \ref{evolve}(c, d). By fixing $r_t$ at $r_c$, Eq.  (\ref{r_m vs r_t}) can successfully fit Fig. \ref{zm vs rm}. This implies that the relation between $z_m$ and $r_m$ is mostly dictated by the geometric constraints from the contact angle at both ends of the water column.

As shown in Fig. \ref{H_star_vs_R}, the theoretical prediction fails to describe the data for $R$ larger than 23 mm. To be precise, the correlation turns negative. This turn of events involves the water jet that shoots up after dropping a stone into a pool.
When $R$ is large, the amount of water that needs to be squeezed out of the liquid column during stage II is also large. This rapid transfer of water 
creates a cavity that is later compressed by hydrostatic pressure, which leads to the formation of a jet.
 After the liquid jet spouts upward and hits the cylinder, it bounces back and carries more water than it brought in. This reduces $r_c$ to a second critical radius that we denote by $r_{c2}<r_c$ whose relation with $R$ and $H^\star$ can be found in Fig. \ref{rc_vs_R}.

It is worth checking the properties often associated with the pinch-off phenomenon because, rather than self-induced, it appears to be hastened by the incoming flow field from the shrinking motion in stage II. The first characteristic we examine is self-similarity by mapping out the neck's contours at successive times leading up to the pinch-off in Fig. \ref{similar} (a). As can be found in Fig. \ref{similar} (b), they fail to converge to a master curve after rescaling \cite{rescale}. Nevertheless, the shrinkage of the neck radius in Fig. \ref{scalingrelation} still obeys the same power law $r_m\propto \tau^{2/3}$  as in ordinary water column \cite{rescale} where $\tau \equiv t_c -t$ and $t_c$ is the pinch-off time.

\section{Conclusion and discussions}
The liquid column formed under a cylinder being lifted out of water is ubiquitous, for instance, in the lapping of dogs, when wild geese take off from a lake, or when we walk through a puddle. Supported by experiments and theoretical models, we elucidate the physics behind the three stages of this phenomenon, which are distinct by different liquid column shapes and dynamics. 

Initially, the system is static and reversible, meaning that the shape of the column, although varying as the height increases, will revert to its former shape when we lower the cylinder.  This stage is characterized by $r_{t}=R$. Reminiscent of the stretching of soap films or bubbles, the shape of the column can be determined by minimizing the total energy consisting of the surface and gravitational energies. Moreover, a critical height $H^\star$ can be deduced when the minimization process fails to produce a static solution, which implies the usher in of an unstable behavior.  

Stage II is characterized by an irreversible contraction of $r_{t}$ at the fixed height $H^\star$. A simple model is proposed to successfully describe the dynamics of $r_{t}$ as well as explain why  $r_{t}$ and the neck radius $r_{m}$ shrink with an equal rate whose magnitude speeds up with time. 

The shrinkage of $r_{t}$ ceases at $r_c$ during the final stage III when $r_{m}\propto \tau^{2/3}$ where $\tau\equiv t_p-t$ and $t_p$ is the pinch-off time. 
We provide a minimal model that can capture the relation between $r_{m}$ and $z_{m}$ that decrease together with time.  Note that most systems manifesting the pinch-off phenomenon exhibit self-similarity ascribed to memory loss on the boundary conditions. It turns out that this property is absent in the liquid column. We suspect the reason is that information of $R$ can be smuggled through the shrinking distance $R-r_c$ of $r_t$ by the kinetic energy carried over from stage II to III.  The plausibility of this scenario is strengthened by its ability to explain another complex behavior accompanying the use of a large cylinder. 

The amount of water squeezed from the column into the pool in stage II is roughly $\pi(R^2-r_c^2)H^\star$. Normally $r_c$ increases with an increasing $R$, according to Fig. \ref{H_star_vs_R}.  However, the trend is reversed, and $r_c$ decreases for $R$ bigger than roughly 13 mm, which implies an increase in liquid transfer. We believe that is responsible for the bulge of water climbing up the column from the pool 
 and interrupting the pinch-off by spreading open the neck when a large cylinder is used. This visible clump of water then bounces off from the cylinder and plunges back into the pool. Afterward, the delayed process of pinch-off resumes. This episode reminds us of the jet made by a dropping stone.
However, the squirting action in our experiment is without the creation and collapse of an air cavity \cite{jet}. We speculate that the jet is made possible by the pressure wave created when the large quantity of water squirted into the pool bounces off from the bottom of the container.  

Apart from clarifying the mechanism that drives the change of morphology and deriving analytic expressions for the critical height and upper radius for the liquid column when transiting between different stages, we can foresee a potential application to the bipedal microrobot \cite{robot}. One key challenge in legged robotic technologies is to ensure a robust and reliable operation in terrains with unpredictable irregularities and disturbances. It is conceivable that the possibility of treading a puddle is unavoidable. Detailed knowledge, such as how much drag force from the liquid column the microrobot is experiencing on one leg in real-time, is crucial to gauge the force the other leg needs to exert to maintain its balance.    

\begin{acknowledgments}
We are grateful to Ting-Heng Hsieh and Wei-Chih Li for useful discussions, and Zhen-Man Tian for technical assistance. 
Financial support from the National Science and Technology Council in Taiwan under Grant No. 111-2112-M007-025 and No. 112-2112-M007-015 is acknowledged.
\end{acknowledgments}

\end{document}